\documentclass{article}
\usepackage{times}
\usepackage{spconf,amsmath,graphicx,xcolor}
\usepackage{multirow,multicol,makecell,arydshln}

\usepackage{pifont}
\newcommand{\cmark}{\ding{51}}%
\newcommand{\xmark}{\ding{55}}%
\usepackage{hyperref}

\title{Crossmodal ASR Error Correction with Discrete Speech Units}
\name{Yuanchao Li, Pinzhen Chen, Peter Bell, Catherine Lai}
\address{University of Edinburgh, UK}

\begin{document}
%
\maketitle
\begin{abstract}
ASR remains unsatisfactory in scenarios where the speaking style diverges from that used to train ASR systems, resulting in erroneous transcripts. To address this, ASR Error Correction (AEC), a post-ASR processing approach, is required. In this work, we tackle an understudied issue: the \textsc{Low-Resource Out-of-Domain} (LROOD) problem, by investigating crossmodal AEC on very limited downstream data with 1-best hypothesis transcription. We explore pre-training and fine-tuning strategies and uncover an ASR domain discrepancy phenomenon, shedding light on appropriate training schemes for LROOD data. Moreover, we propose the incorporation of discrete speech units to align with and enhance the word embeddings for improving AEC quality. Results from multiple corpora and several evaluation metrics demonstrate the feasibility and efficacy of our proposed AEC approach on LROOD data as well as its generalizability and superiority on large-scale data. Finally, a study on speech emotion recognition confirms that our model produces ASR error-robust transcripts suitable for downstream applications.
\end{abstract}
\section{Introduction}
\label{sec:intro}

Over the past decade, the field of Automatic Speech Recognition (ASR) has made significant progress, driven by improvements in computing resources and training schemes, as well as the availability of data. In particular, speech foundation models have demonstrated excellent performance \cite{baevski2020wav2vec,hsu2021hubert,radford2023robust}, pushing the boundaries of ASR performance to new heights. In addition, their learned representations have proven valuable not only for ASR but also for various downstream applications and scenarios \cite{li2023exploration,yang2023device,gong2023whisper}.

However, in some situations, the speech data does not have much in common with the data used to train the ASR systems, resulting in an out-of-domain problem. For example, the acoustic characteristics of emotional, pathological, or children's speech could contain irregular, disfluent, and ambiguous patterns, which are different from and unexpected in the speech found in audiobooks or general speaking scenarios \cite{li2023asr,wu2023self,Dutta2022ChallengesRI}. As a result, when ASR is applied to these downstream tasks (e.g., emotion or depression detection), the transcription is of poor quality and difficult to use \cite{wu2023self,li2022fusing}.

To further improve the performance of ASR, ASR Error Correction (AEC) approaches have been proposed, allowing for post-processing without modifying the acoustic model. Traditionally, researchers trained an external language model to be incorporated into the ASR system for re-scoring \cite{tanaka2018neural}. More recently, there has been a trend toward using generative error correction via Large Language Models (LLMs) \cite{yang2023generative}, replacing traditional language models. Additionally, end-to-end AEC, which maps erroneous transcripts to ground-truth text using a Sequence-to-Sequence (S2S) approach, has become prevalent in scenarios where ASR is treated as a black box \cite{mani2020asr,liao2023improving}. Furthermore, some work has used both acoustic information and ASR hypotheses as input instead of text-only data, achieving crossmodal AEC \cite{lin2023multi,du2022cross,radhakrishnan2023whispering,chen2024s}. 

Despite these advances, AEC is still a challenging task, especially for \textit{Low-Resource Out-of-Domain (LROOD) data}. Therefore, we explore this relatively unexplored aspect, aiming to provide a comprehensive analysis, which in turn provides a better understanding of AEC, as well as to improve the crossmodal AEC using Discrete Speech Units (DSUs). The exploration steps with respective research problems and hypotheses are as follows.

\textbf{\textit{1)}} While S2S models have been established for AEC, research on LROOD scenarios is limited. \textit{Many previous studies were performed on the same large corpus without considering the LROOD problem \cite{zhang2021end,tanaka2021cross}, leaving challenges remain such as determining effective Pre-Training (PT) and Fine-Tuning (FT) strategies with LROOD data}. Therefore, we compare AEC performance with and without PT or FT on LROOD data using an S2S model.

\textbf{\textit{2)}} None of the prior works has considered the characteristics of the ASR models that are the source of transcript generation. Moreover, although some studies have used data augmentation to produce more erroneous sentences for LROOD downstream corpora for AEC training, we argue that such arbitrary augmentation is unreliable because the error patterns of the augmented data differ from the original ASR errors. \textit{We hypothesize that different ASR models may produce distinct patterns of ASR errors (e.g., some may have more insertions, substitutions, or deletions, and some may remove or retain disfluencies), which requires that the AEC model be trained for corresponding ASR domain errors}. Thus, through a comparative analysis using transcripts obtained from different ASR models with nearly the same WER, we investigate this issue and refer to it as \textsc{asr domain discrepancy}.

\textbf{\textit{3)}} Acoustic information has proven useful for crossmodal AEC \cite{lin2023multi,radhakrishnan2023whispering}, but \textit{it is not always possible to acquire audio sources for the PT stage (e.g., due to privacy or other ethical issues)}. Therefore, determining how to better incorporate audio features and which acoustic features are useful remains an open question, considering high-WER speech usually contains low-quality audio that can introduce distortions into the crossmodal training. To address this, we improve crossmodal AEC by incorporating DSUs only in the FT stage, representing a resource-efficient and effort-saving approach.

\textbf{\textit{4)}} \textit{Very few studies have applied corrected ASR transcripts to downstream tasks to evaluate AEC extrinsically}. Hence, we conduct Speech Emotion Recognition (SER) using the transcripts corrected by our proposed AEC approach, validating its potential for downstream applications.

\noindent \textbf{Related Work}: To our knowledge, we are the first to address the LROOD problem in AEC. The closest works are \cite{lin2023multi, chen2024s, zhang2021end, tanaka2021cross}, which fused audio into the language model as crossmodal AEC. They either fine-tuned models on large downstream data or used the same corpus with audio in all phases of model training. This work, however, proposes DSU fusion in FT only and provides insights for tasks that require high-quality transcripts yet are constrained by limited resources.

\section{Corpora, ASR Models, and Metrics}
\label{sec:corpora}
Two corpora are primarily used in this work: Common Voice \cite{ardila2019common} and IEMOCAP \cite{busso2008iemocap}. Common Voice is one of the largest publicly available multilingual and diverse speech datasets. We adopt its 13.0 English version, using 150k utterances from the training set and the entire test set, bringing the total number to 166k. IEMOCAP is an acted corpus consisting of dyadic sessions in which actors perform improvisations or scripted scenarios specifically selected to elicit emotional expression. We follow prior research \cite{li2023asr}, using four basic emotions with 5,500 utterances whose transcripts are not empty.

CMU-MOSI \cite{zadeh2016multimodal} and MSP-Podcast \cite{lotfian2017building} are also used for validation, with results presented, but detailed analysis is omitted due to limited space. CMU-MOSI contains only 2,199 samples, which is approximately half the size of IEMOCAP. We randomly select 1,800 samples to fine-tune our AEC model and the remaining samples for testing. For MSP-Podcast, we adopt its Odyssey 2024 emotion challenge version, which contains a training set of 68,119 samples and a test set of 19,815 samples, making it approximately 16 times larger than IEMOCAP. These two corpora are used for: \textbf{\textit{1})} confirming the generalizability of our approach and further proving its effectiveness; and \textbf{\textit{2)}} investigating the impact of data size (i.e., how our performance varies with different amounts of FT data).

For the ASR models, we use \textit{Wav2Vec 2.0} (\textit{W2V2}) \cite{baevski2020wav2vec} in its \textit{base-960h} version, a \textit{Conformer} model (\textit{CONF}) \cite{gulati2020conformer} from \textit{ESPnet} \cite{watanabe2018espnet}, and the \textit{Whisper} model \cite{radford2023robust} in its \textit{tiny.en} version. We combine the transcripts of \textit{W2V2}, \textit{CONF}, and the ground truth, resulting in a mixture of different system error types that mimic the transcript of a random ASR system (\textit{Random}). This mixture is then compared with the \textit{Whisper}-based transcript with a comparable WER to investigate the ASR domain discrepancy problem in AEC (see Sec.~\ref{sec:domain}).

\textit{Whisper} is run on all four corpora, yielding satisfactory WERs (see Table~\ref{tab:wer}) for the following reasons: \textbf{\textit{1)}} the WERs indicate that the corpora fall outside the domain used for training \textit{Whisper}, aligning with our goal to study the OOD problem; \textbf{\textit{2)}} the WER of Common Voice (for PT) is close to the others (for FT), ensuring the error ratios are consistent in PT and FT. Otherwise in PT, too many errors can result in a serious over-correction problem in subsequent FT and inference phases (which generally occurs in OOD scenarios, as we have observed in our experiments), while too few errors may lead to insufficient learning of error-gold pairs. This setting was usually ignored in the literature, where many studies trained the AEC model on Librispeech with less than 10\% WER \cite{guo2019spelling}, thereby hindering their generalizability. Subsequently, to discover the ASR domain discrepancy problem, \textit{which is a new concept presented by this work}, we create the transcript of IEMOCAP from the \textit{Random} model for comparison, which has almost the same WER as that from \textit{Whisper}. As said, we omit experimental details on CMU-MOSI and MSP-Podcast.

\begin{table}[ht]
\centering
\caption{WERs (\%) of the ASR transcripts.}
\begin{tabular}{lll}
\hline
\textbf{ASR Model} & \textbf{Corpus} & \textbf{WER} \\ \hline
\multirow{2}{*}{\textit{Whisper}} & Common Voice & 19.11 \\
 & IEMOCAP & 17.18 \\
 & CMU-MOSI & 17.84 \\
 & MSP-Podcast & 17.65 \\ \hline
\textit{Random} & IEMOCAP & 17.12 \\ \hline
\end{tabular}
\label{tab:wer}
\end{table}

In the following experiments, we employ three metrics to evaluate AEC performance: WER, BLEU, and GLEU. Unlike WER, which examines individual words, BLEU considers n-grams as units for further validating the quality of corrected transcripts \cite{mani2020asr}. Additionally, GLEU \cite{wu2016google} is designed with a specific focus on per-sentence reward objectives, addressing certain limitations associated with applying BLEU to individual sentences. We believe that BLEU and GLEU analyze a broader word span containing context for downstream tasks to infer syntactic and semantic information. Utilizing all three metrics, we aim to comprehensively assess AEC quality from different perspectives.

\section{Approach, Experiments, and Results}

\begin{figure*}[ht]
  \centering
  \includegraphics[width=0.88\textwidth]{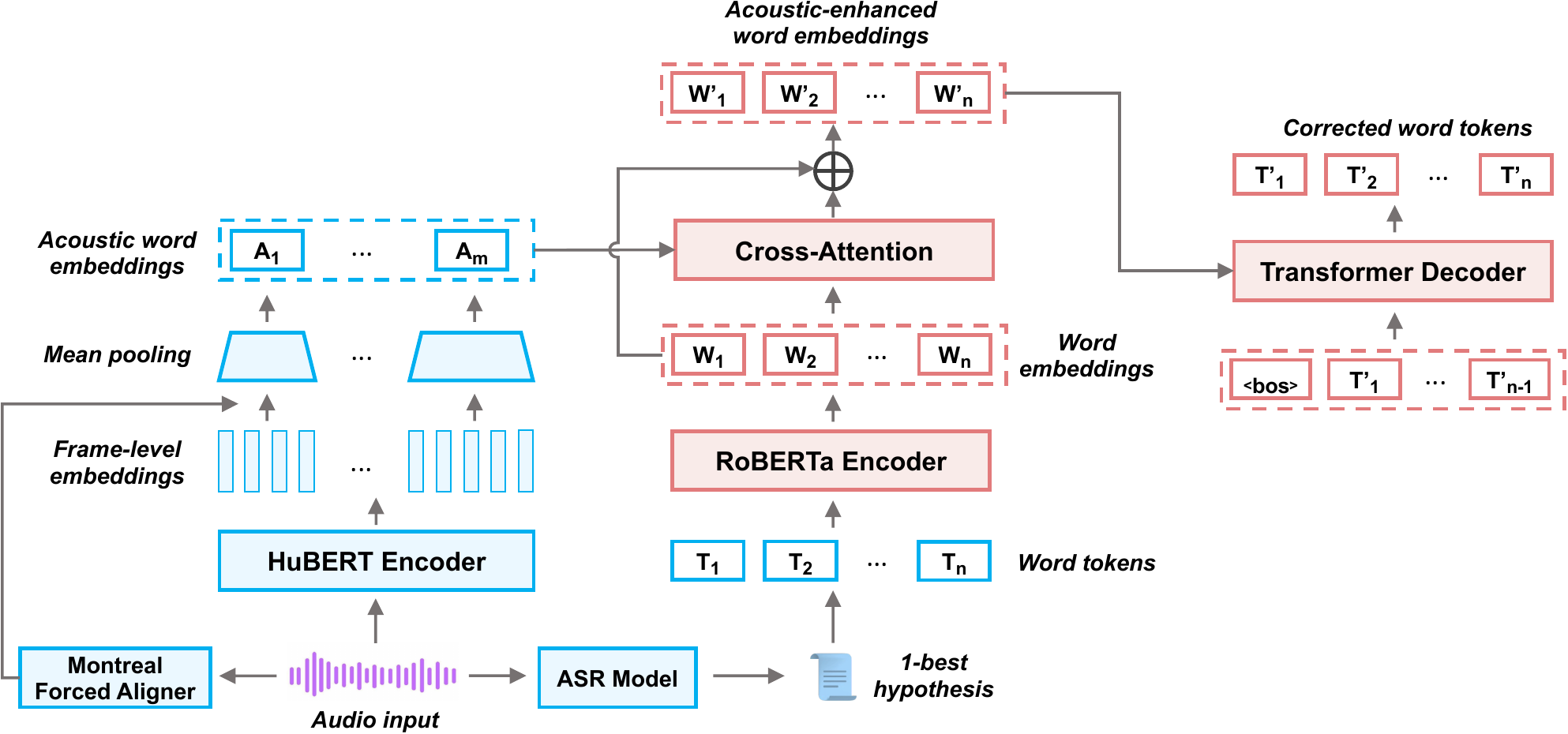}
  \caption{Architecture of our crossmodal AEC with discrete speech units. (\textcolor{pink}{Pink}: trainable; \textcolor{cyan}{Blue}: frozen).}
  \label{fig:model}
\end{figure*}

The architecture of our proposed AEC approach is shown in Fig.~\ref{fig:model}. On one path, the audio input is transcribed by the ASR model into a textual transcript, which is then tokenized for the \textit{RoBERTa-base} encoder to generate word embeddings. On the other path, the \textit{HuBERT} encoder produces Self-Supervised Representations (SSR) from the audio input, followed by mean pooling to generate SSR-based Acoustic Word Embeddings (AWEs) as DSUs. Note that the Montreal Forced Aligner \cite{mcauliffe2017montreal} was used on the ASR transcripts beforehand to determine the word boundaries for mean pooling. Next, cross-attention aligns the AWEs and word embeddings to obtain the acoustic-enhanced word embeddings for the Transformer decoder to produce corrected word tokens. Our motivation for using this architecture is to highlight our effectiveness even with the most basic components.

\subsection{Crossmodal AEC on LROOD Data}
\label{sec:ood}
We first investigate PT and FT without incorporating audio information, revealing the ASR domain discrepancy problem mentioned earlier. Subsequently, we incorporate different acoustic features and propose the use of DSUs for better audio-text alignment to generate corrected word tokens.

\subsubsection{Pre-Training \& Fine-Tuning}
To pre-train the AEC model, 166k samples from Common Voice were recognized by \textit{Whisper}, with 1,000 random samples held out as the test set, and the rest for training and validation with an 80\%-20\% split. The training aims to recover the gold transcripts from the ASR output. With a batch size of 256, an initial learning rate of 1e-5, and the Adam optimizer, we train the model for 30 epochs using cross-entropy loss and select the best checkpoint based on WER as the evaluation metric. Decoding is performed using beam search with a size of 5. The training framework is adopted from \cite{chen2023exploring}. Performance on the test set is shown in Table~\ref{tab:encoder}.

\begin{table}[ht]
\centering
\caption{AEC Performance on the test set of Common Voice.}
\scalebox{0.93}{
\begin{tabular}{lcccc}
\hline
\textbf{Model} & \textbf{WER} & \textbf{BLEU} & \textbf{GLEU} \\
\hline
\textit{Original ASR transcript} & 19.30 & 70.56 & 71.24 \\ 
\textit{Best checkpoint} & 18.19 & 72.14 & 72.45 \\
\hline
\end{tabular}
}
\label{tab:encoder}
\end{table}

Furthermore, to prevent the over-correction problem, we continue training this saved checkpoint on TED transcriptions \cite{cettolo-etal-2012-wit3} to learn to copy the gold transcripts (i.e., ground truth $\rightarrow$ ground truth) and potentially enhance its domain robustness. This continue-training lasts for two epochs, ensuring it does not overfit while maintaining correction stability. We save the checkpoint\footnote{\href{https://github.com/yc-li20/Crossmodal_AEC}{https://github.com/yc-li20/Crossmodal\_AEC}} as the base model for subsequent experiments.

Next, we fine-tune this model on the training set of IEMOCAP for 40 epochs with a batch size of 64, an initial learning rate of 2e-5 (excluding the parameters of ``bias'' and ``LayerNorm.weight''), an epsilon of 1e-8, and the Adam optimizer. Following the standard five-fold split of IEMOCAP, the FT is performed five times (each time using four folds for FT and one for testing), and the final performance is reported based on the transcript composed of the corrected results obtained from five instances. The details of FT on the other two corpora are omitted yet the final results will be presented in Sec.~\ref{sec:general}.

\begin{table}[ht]
\centering
\caption{Comparison results on IEMOCAP of w/ and w/o pre-training or fine-tuning.}
\scalebox{0.93}{
\begin{tabular}{ccccc}
\hline
\textbf{PT} & \textbf{FT} &  \textbf{WER$\downarrow$} & \textbf{BLEU$\uparrow$} & \textbf{GLEU$\uparrow$}  \\ \hline
\multicolumn{2}{c}{\textit{Original ASR transcript}} & 17.18 & 76.56 & 75.29 \\ \hdashline
\cmark & \xmark & 17.14 & 76.61 & 75.34 \\
\xmark & \cmark & 17.08 & 77.01 & 75.52 \\
\cmark & \cmark & \textbf{16.40} & \textbf{78.00} & \textbf{76.58} \\ \hline
\end{tabular}
}
\label{tab:pre_con}
\end{table}

Table~\ref{tab:pre_con} shows that without FT, a pre-trained model cannot perform well. The improvement is hardly noticeable on IEMOCAP, whereas the improvement is significant on the test set of Common Voice (Table~\ref{tab:encoder}), despite their original ASR transcripts being of similar quality (Table~\ref{tab:wer}). This is likely due to the domain discrepancy between Common Voice and IEMOCAP, which results in the model pre-trained on the former being unable to recognize some erroneous OOD words in the latter. However, even without PT, the model can still improve transcript quality after FT\footnote{Technically, since there is no PT on Common Voice, it is not appropriate to use the term ``FT'' as the model is directly trained on IEMOCAP. However, we keep ``FT'' here for consistency.} on LROOD data.

Our best result comes from using both PT and FT, which indicates that the capacity learned during PT is activated and enhanced by FT. This combination well alleviates the LROOD problem.

\subsubsection{ASR Domain Discrepancy Problem}
\label{sec:domain}
To study the impact of ASR domain discrepancy, we conduct experiments by FT the AEC model on the output from another ASR system. Specifically, we use the transcript generated by \textit{Random} and compare it to the transcript generated by \textit{Whisper}. The results are presented in Table~\ref{tab:dis}.

We can observe that without PT on the transcript of Common Voice (which is generated by \textit{Whisper}), the difference in the metric values remains small after FT\footnotemark[2], compared to their original ASR transcripts. However, this pattern disappears with PT, as the transcript quality from \textit{Whisper} becomes better than that from \textit{Random}, highlighting the detrimental impact of ASR domain discrepancy. This phenomenon suggests that to correct transcripts from an ASR model, it is crucial to use the same ASR model as that used in PT (i.e., to continue using the same ASR model in both PT and FT). Nevertheless, the transcript quality from \textit{Random} still improves, indicating that PT on a large corpus, even if its transcript is from a different ASR model, is still indispensable in LROOD scenarios.

\begin{table}[ht]
\centering
\caption{Comparison results on IEMOCAP of fine-tuning on transcript generated by different ASR models.}
\scalebox{0.93}{
\begin{tabular}{lccc}
\hline
\textbf{ASR Model} &  \textbf{WER$\downarrow$} & \textbf{BLEU$\uparrow$} & \textbf{GLEU$\uparrow$}  \\ \hline
\multicolumn{4}{l}{Original ASR transcript} \\
\textit{Whisper} & 17.18 & 76.56 & 75.29 \\
\textit{Random} & 17.12 & 76.64 & 75.38 \\ \hdashline
\multicolumn{4}{l}{\xmark pre-training} \\
\textit{Whisper} & 17.08 & 77.01 & 75.52  \\
\textit{Random} & 17.03 & 77.08 & 75.61 \\ \hdashline
\multicolumn{4}{l}{\cmark pre-training} \\
\textit{Whisper} & 16.40 & 78.00 & 76.58 \\
\textit{Random} & 16.54 & 77.57 & 76.42 \\ \hline
\end{tabular}}
\label{tab:dis}
\end{table}

\subsubsection{Incorporation of Discrete Speech Units}
\label{sec:dsu}
So far, we have investigated how PT and FT contribute to text-only S2S AEC. To further improve the quality of error correction, we study the incorporation of acoustic information. Previous studies usually incorporated acoustic information in all stages—PT, FT, and testing—and only utilized continuous features such as Mel-spectrogram or raw SSR \cite{lin2023multi,du2022cross}. However, we argue that these practices do not apply to LROOD scenarios for the following reasons:

\textbf{\textit{1)}} The audio source of large-scale PT data is not always accessible due to privacy or other ethical concerns. \textbf{\textit{2)}} The high-WER OOD speech usually contains low-quality audio that can introduce acoustic distortions (e.g., prosody variation or noise) into crossmodal training. \textbf{\textit{3)}} It is challenging to align discrete word embeddings with continuous audio features. To this end, we propose to discretize the audio features to create DSUs and avoid incorporating such acoustic information in PT, making it a resource-efficient and effort-saving approach.

We utilize AWEs, which are fixed-dimensional vectors representing variable-length spoken word segments as DSUs. These vectors map acoustic features extracted from audio signals to vectors, where similar words or linguistic units have similar embeddings in the vector space \cite{maas2012word,levin2013fixed}. AWEs can capture information about phonetics and other acoustic aspects of speech, offering promising potential for word discrimination \cite{matusevych2020analyzing}.

Following recent studies on the analysis of AWEs from self-supervised speech models, we use SSR from \textit{HuBERT} with mean pooling followed by forced alignment to find the word boundary, as this practice has been shown competitive with the state of the art on English AWEs \cite{sanabria2023analyzing,saliba2024layer}. On the other hand, we also use Mel-spectrogram and continuous raw SSR for comparison. After a layer-wise analysis (omitted due to space), we use AWEs from \textit{HuBERT} layer 7 and raw SSR from \textit{HuBERT} layer 8 as they performed the best among all layers, respectively. This aligns with a previous finding that \textit{HuBERT} encodes the most word information between the middle layer and the last layer \cite{pasad2023comparative}.

To incorporate the DSUs, we set the maximum sequence length as that of corresponding word embeddings and 0-pad the short sequence. To incorporate continuous features for comparison, we first downsample them to the same sequence length as the word embeddings using a fast Fourier transform. Unlike \textit{HuBERT}, which has the same feature dimension of 768 as \textit{RoBERTa}, we use a feed-forward layer for Mel-spectrogram to expand its dimension to this size. After such pre-processing, we implement cross-attention to align acoustic features with word embeddings:
\begin{align}
A' = Attn(Q_{w},K_{a},V_{a}) = softmax(\frac{Q_{w}K_{a}^T}{\sqrt{d_k}})V_{a}
\end{align}
where $Q_{t}$, $K_{a}$, and $V_{a}$ represent the respective matrix for query (word embeddings), key (acoustic features), and value (acoustic features), $d_{k}$ is the size of a key vector, and $A'$ is the word-aligned acoustic features. Next, we add $A'$ and $W$ for the Transformer decoder with optimizable parameters $\theta_{T}$ to generate a corrected version $W'$:
\begin{align}
W'=\arg\!\max_{W}P(W| addition(A', W);\theta_{T})
\end{align}
The results of fusing acoustic features are shown in Table~\ref{tab:summary} with previous experimental results included for comparison.

\begin{table}[ht]
\centering
\caption{Result summary on IEMOCAP.}
\scalebox{0.93}{
\begin{tabular}{lccc}
\hline
\textbf{Model} & \textbf{WER$\downarrow$} & \textbf{BLEU$\uparrow$} & \textbf{GLEU$\uparrow$}  \\ \hline
\textit{Original ASR transcript} & 17.18 & 76.56 & 75.29 \\ \hdashline
\textit{PT} & 17.14 & 76.61 & 75.34 \\
\textit{FT} & 17.08 & 77.01 & 75.52 \\
\textit{PT+FT} & 16.40 & 78.00 & 76.58 \\
\textit{PT+FT+Mel-spec} & 17.36 & 76.82 & 75.48 \\
\textit{PT+FT+HuBERT SSR} & 16.20 & 78.01 & 76.71 \\
\textit{PT+FT+HuBERT AWEs} & \textbf{16.07} & \textbf{78.22} & \textbf{76.96} \\ \hline
\end{tabular}
}
\label{tab:summary}
\end{table}

We note that \textbf{\textit{1)}} compared with other acoustic features, \textit{HuBERT} AWEs provide the best results across all metrics. This verifies our hypothesis that \textit{DSUs align more easily with word embeddings than continuous acoustic features}. \textbf{\textit{2)}} The inclusion of Mel-spec worsens WER rather than improves it, which contrasts with findings in \cite{lin2023multi,zhang2021end,tanaka2021cross}. This phenomenon is reasonable and consistent with discussions in Sec.~\ref{sec:dsu}: \textit{i)} IEMOCAP being emotional speech, contains intense prosody variation, making it challenging to encode phonetic information from Mel-spec; \textit{ii)} the small-size data for FT (4.4k samples with an average duration of 5 seconds) hinders the model from sufficiently learning linguistic information from Mel-spec, representing a low-resource scenario; \textit{iii)} our incorporation of audio features only happens during FT and testing, causing Mel-spec to struggle to provide sufficient information to word embeddings. In contrast, \cite{lin2023multi,zhang2021end,tanaka2021cross} conducted all training phases using the same large corpus, making their findings inapplicable to LROOD scenarios. \textbf{\textit{3)}} Interestingly, despite that Mel-spec worsens WER compared to the original ASR transcript and PT, BLEU and GLEU record improvement. This is likely because the corrected texts are more fluent and structurally correct with respect to the reference (favourable for BLEU and GLEU), while still containing word-level mistakes captured by WER. This demonstrates the contribution of audio to high-level linguistic information, which corroborates our later finding in SER (Sec.~\ref{sec:ser}).

\subsubsection{Evaluation on Additional Corpora}
\label{sec:general}
As mentioned before, we test the performance of our proposed approach on two more corpora: CMU-MOSI and MSP-Podcast, to verify its generalizability. The results are shown in Table~\ref{tab:mosi} and \ref{tab:msp}. All experimental settings remain the same, while several non-optimal models are omitted for brevity.

It can be noted that \textbf{\textit{1)}} PT fails to provide better results than the original ASR transcript on CMU-MOSI, whereas the performance improvement is significant on MSP-Podcast. This phenomenon is plausible due to the OOD problem: CMU-MOSI consists of monologue speech with opinions on specific topics (mainly about movies), containing a high proportion of OOD words, making the PT model trained on Common Voice less effective. In contrast, MSP-Podcast consists of natural, real-life speech recorded in podcast settings, sharing more linguistic similarities with Common Voice. \textbf{\textit{2)}} Both FT and the incorporation of DSUs bring performance improvements on CMU-MOSI, despite PT not being effective and the FT data being extremely limited at only 1,800 samples. Since the data size and domain similarity of IEMOCAP are between those of CMU-MOSI and MSP-Podcast, its performance improvement also falls in between (Table~\ref{tab:summary}). Furthermore, the performance improvement is even more significant on MSP-Podcast, indicating that the more data available for FT, the better the performance. These findings demonstrate the efficacy of our approach in LROOD scenarios and also highlight its generalizability and potential across various scenarios.

\begin{table}[ht]
\centering
\caption{Result summary on CMU-MOSI.}
\scalebox{0.93}{
\begin{tabular}{lccc}
\hline
\textbf{Model} & \textbf{WER$\downarrow$} & \textbf{BLEU$\uparrow$} & \textbf{GLEU$\uparrow$}  \\ \hline
\textit{Original ASR transcript} & 17.84 & 72.82 & 72.17 \\ \hdashline
\textit{PT} & 17.88 & 72.80 & 72.16 \\
\textit{PT+FT} & 17.65 & 73.31 & 72.63 \\
\textit{PT+FT+HuBERT AWEs} & \textbf{17.22} & \textbf{73.98} & \textbf{73.01} \\ \hline
\end{tabular}
}
\vspace{-7pt}
\label{tab:mosi}
\end{table}

\begin{table}[ht]
\centering
\caption{Result summary on MSP-Podcast.}
\scalebox{0.93}{
\begin{tabular}{lccc}
\hline
\textbf{Model} & \textbf{WER$\downarrow$} & \textbf{BLEU$\uparrow$} & \textbf{GLEU$\uparrow$}  \\ \hline
\textit{Original ASR transcript} & 17.65 & 81.32 & 78.02 \\ \hdashline
\textit{PT} & 16.23 & 82.59 & 79.14 \\
\textit{PT+FT} & 14.73 & 83.16 & 80.84 \\
\textit{PT+FT+HuBERT AWEs} & \textbf{13.89} & \textbf{83.64} & \textbf{81.80} \\ \hline
\end{tabular}
}
\label{tab:msp}
\end{table}

\subsubsection{Performance Comparison with Literature}
To confirm the effectiveness of our approach, we compare it with the literature by adopting the following baselines:

\textbf{\textit{1)}} Crossmodal AEC using continuous acoustic information: Mel-spectrogram \cite{zhang2021end}

\textbf{\textit{2)}} Crossmodal AEC using continuous acoustic information: raw self-supervised representations \cite{lin2023multi}.

\textbf{\textit{3)}} Generative AEC using an LLM with 1-best ASR hypothesis and Alpaca prompt \cite{radhakrishnan2023whispering}.

\textbf{\textit{4)}} Generative AEC using an LLM with N-best ASR hypothesis and Alpaca prompt \cite{radhakrishnan2023whispering}.

\textbf{\textit{5)}} Generative AEC using an LLM with 1-best ASR hypothesis and Task-Activating prompt \cite{yang2023generative}.

\textbf{\textit{6)}} Generative AEC using an LLM with N-best ASR hypothesis and Task-Activating prompt \cite{yang2023generative}.

\begin{figure}[ht!]
  \centering
  \includegraphics[width=0.515\textwidth]{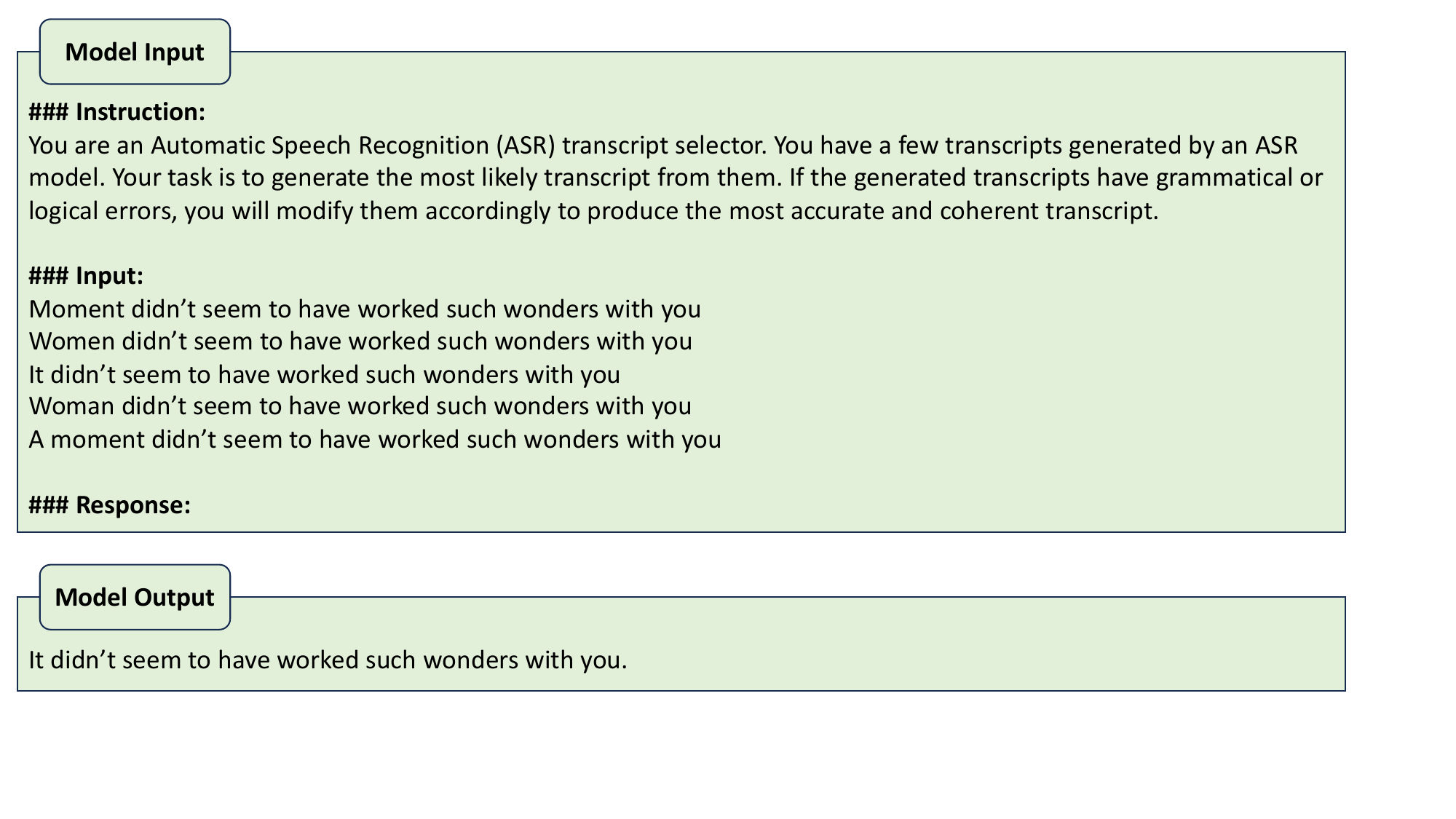}  
  \includegraphics[width=0.520\textwidth]{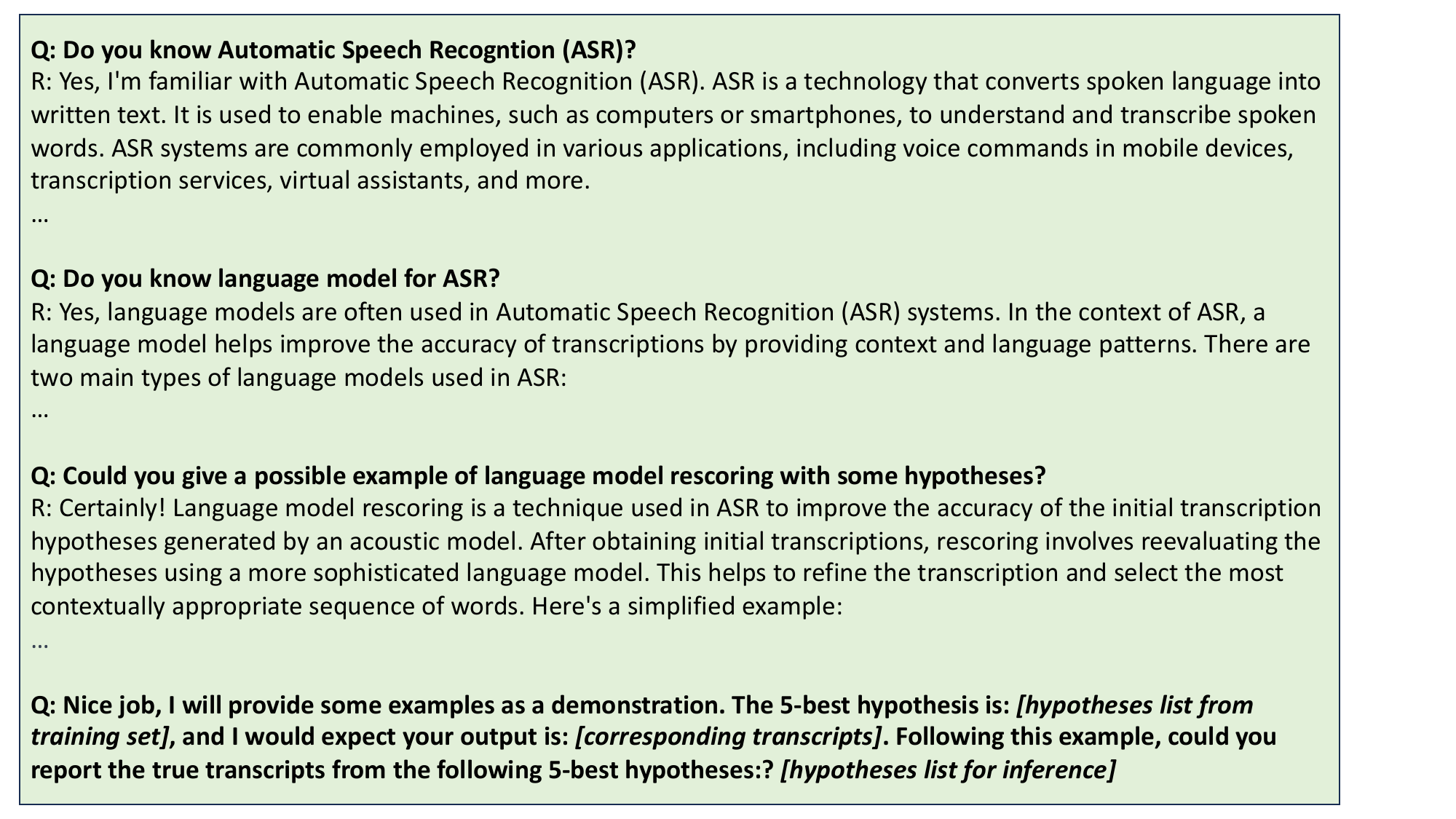}
  \caption{An illustration of the Alpaca prompt (upper) and Task-Activating prompt (below) used in this work.}
  \label{fig:prompt}
\end{figure}

Since the comparisons with \textbf{\textit{1)}} and \textbf{\textit{2)}} have already been presented in Table~\ref{tab:summary} and discussed, we omit them here. For the remaining comparisons, we attempt the Alpaca prompt \cite{taori2023stanford} and Task-Activating (TA) prompt \cite{yang2023generative} using \textit{InstructGPT} on both 1-best and 5-best hypotheses. Fig.~\ref{fig:prompt} illustrates how the Alpaca prompt and TA prompt are used. The results are presented in Table~\ref{tab:comparison}.

\begin{table}[ht]
\centering
\caption{Performance comparison with generative AEC approaches.}
\scalebox{0.93}{
\begin{tabular}{lccc}
\hline
\textbf{Model} & \textbf{WER$\downarrow$} & \textbf{BLEU$\uparrow$} & \textbf{GLEU$\uparrow$}  \\ \hline
\textit{Original ASR transcript} & 17.18 & 76.56 & 75.29 \\
\textit{Our full model} & \textbf{16.07} & \textbf{78.22} & \textbf{76.96} \\ \hdashline
\textit{\textbf{3)} Alpaca prompt\textsubscript{1-best}} & 17.18 & 76.56 & 75.29 \\
\textit{\textbf{4)} Alpaca prompt\textsubscript{5-best}} & 17.01 & 76.97 & 75.44 \\
\textit{\textbf{5)} TA prompt\textsubscript{1-best}} & 17.18 & 76.57 & 75.30 \\
\textit{\textbf{6)} TA prompt\textsubscript{5-best}} & 16.62 & 77.99 & 75.98 \\ \hline
\end{tabular}
}
\label{tab:comparison}
\end{table}

From the comparison results, it can be observed that: the generative AEC approaches underperform our S2S crossmodal AEC approach, particularly as the 1-best hypothesis shows hardly any difference compared to the original ASR transcript, confirming our effectiveness for scenarios where only the 1-best hypothesis is available.

\section{AEC for Downstream Use -- SER}
\label{sec:ser}
To verify the quality and usability of our AEC approaches in downstream applications, we compare SER performances using the corrected transcript and the original ASR transcript.

Following the same training scheme as \cite{li2023asr}, we train the SER model on the ground-truth transcript of the IEMOCAP training set and evaluate its performance on the ASR transcript of the test set, employing five-fold cross-validation. Textual features are extracted using \textit{BERT}. The SER model consists of two bidirectional LSTM layers (hidden state: 32), a self-attention layer (hidden state: 64, heads: 16), a dense layer (hidden state: 64) with ReLU activation, and an output layer with Softmax activation. We use the AdamW optimizer with a learning rate of 1e-4 and weight decay of 1e-5 and a batch size of 64. Training is performed for 150 epochs, and the reported results are the best Unweighted Accuracy (UA) achieved. The only difference from \cite{li2023asr} is that they used the pooler output from \textit{BERT}, while we use hidden states.

\begin{table}[ht]
\centering
\caption{Comparison results of SER performance.}
\scalebox{0.93}{
\begin{tabular}{lccc|c}
\hline
\textbf{Transcript} & \textbf{WER$\downarrow$} & \textbf{BLEU$\uparrow$} & \textbf{GLEU$\uparrow$} & \textbf{UA$\uparrow$}  \\ \hline
\textit{Original} & 17.18 & 76.56 & 75.29 & 60.92 \\
\textit{Corrected} & 16.07 & 78.22 & 76.96 & \textbf{61.82} \\ \hline
\end{tabular}
}
\label{tab:ser}
\end{table}

As expected, SER performance can be improved by using the corrected transcript. In \cite{li2023asr}, UA increased from 55.4 to 57.1 (\textit{+1.70}) with WER decreasing from 20 to 15 (\textit{-5.00}). In our case, UA increased from 60.92 to 61.82 (\textit{+0.90}) with WER decreasing from 17.18 to 16.07 (\textit{-1.11}), which represents a more significant improvement. This observation can be attributed to the fact that AEC with DSUs not only reduces WER but potentially does so via preserving syntax and semantics better, leading to higher usability in downstream tasks (resonates with the last finding in Sec.~\ref{sec:dsu}). However, further analysis is needed to understand the nature of ASR errors: where they occur and how they are corrected.

\section{Discussion and Conclusion}
In this paper, we pre-trained an S2S AEC model on large corpora and fine-tuned it on an LROOD corpus with the assistance of DSUs. The results indicate that for AEC on LROOD data, PT, FT, and DSUs are all important. Moreover, the ASR domain discrepancy problem requires attention and should be alleviated by using the same ASR model to generate transcripts in all phases of AEC applications. We compared different acoustic features and verified the superiority of DSUs over continuous features in aligning with word embeddings. A downstream task of SER further demonstrated the improved quality of the corrected transcript, highlighting the applicability of our approach. Additionally, as confirmed by the experiment in Sec.~\ref{sec:general}, our approach is expected to perform better on larger downstream data.

\bibliographystyle{IEEEbib}
\bibliography{refs}

\end{document}